\documentclass[twocolumn,showpacs,preprintnumbers,amsmath,amssymb]{revtex4}%,showkeys
\usepackage{graphicx}% Include figure files
\usepackage{dcolumn}% Align table columns on decimal point
\usepackage{bm}% bold math
\begin{document}

%\begin{CJK*}{GBK}{song} %显示中文

\preprint{}

%of ultracold atomic gas

\title{One-step implementation of entanglement  generation on microwave photons in distant 1D superconducting resonators}

\author{Ming Hua, Ming-Jie Tao,  and Fu-Guo Deng\footnote{Corresponding author:fgdeng@bnu.edu.cn}}

\address{Department of Physics, Applied Optics Beijing Area Major Laboratory, Beijing normal University, Beijing 100875, China}
\date{\today }

\begin{abstract}
We present a scalable quantum-bus-based device for generating the
entanglement on microwave photons (MPs) in distant superconducting
resonators (SRs). Different from the processors in previous works
with some resonators coupled to a superconducting qubit (SQ), our
device is composed of some 1D SRs $r_j$ which are coupled to the
quantum bus (another common resonator $R$) in its different
positions simply, assisted by superconducting quantum interferometer
devices. By using the technique for catching and releasing a MP
state in a 1D SR, it can work as an entanglement generator or a node
in quantum communication.  To demonstrate the performance of this
device, we propose a one-step scheme to generate high-fidelity Bell
states on MPs in two distant SRs. It works in the dispersive regime
of $r_j$ and $R$, which enables us to extend it to generate
high-fidelity multi-Bell states on different resonator pairs
simultaneously.
\end{abstract}

\pacs{03.67.Lx, 03.67.Bg, 85.25.Dq, 42.50.Pq}   \maketitle

Quantum entanglement on photons plays a critical role in quantum
communication \cite{entanglement}. There are some interesting ways
for entanglement generation, including those based on linear optical
elements and the input-output process of single photons assisted by
cavity quantum electrodynamics (QED) in which a two-level atom is
coupled to a single-mode field. By far, some artificial atoms have
also been widely studied for quantum information processing, such as
quantum dots \cite{Xiao}, superconducting qubits
\cite{Blais,Wallraff}, and diamond nitrogen-vacancy centers
\cite{Jelezko}. In recent years, quantum information processing on
microwave photons assisted by a superconducting qubit (SQ)
\cite{Chirolli,Hoi,Adhikari,Neumeier} has attracted much attention
because of its good performance on the extinction of the transmitted
microwave photon (MP) \cite{Shen,Astafiev} and the long-distance
traveling along a superconducting transmission line \cite{Kevin}.

Recently, catching and releasing a MP state in a 1D superconducting
resonator (SR) has been realized in experiments with high fidelity
\cite{Yin}, which gives us another way to generate the entanglement
on MPs for quantum communication with the following processes: (1),
MPs can be caught from the transmission line (TL) to a device based
on 1D SRs; (2), quantum entangling operations are performed on
localized MPs assisted by the circuit QED; (3), MPs are released
from the processor to the TL. The entanglement generation and the
universal quantum gates on the localized MPs in different 1D SRs
have been studied both in experiment \cite{Zakka,Wang} and in theory
\cite{Xiong,Hua1}. For example, in 2011, Wang \emph{et al.}
\cite{Wang} realized the entangled NOON state on two 1D storage SRs
in a device composed of a coupling resonator coupled to two phase
qubits and each of the qubits is coupled to a storage SR, and Hu and
Tian presented a scheme to deterministically generate the entangled
photon pairs in a SR array. In 2012, Strauch \cite{Strauch} proposed
an interesting all-resonant method to control the quantum state of
1D SRs in a processor composed of two 1D SRs coupled to a SQ and
Yang \emph{et al.} \cite{Yang} presented a scheme to generate a
Greenberger-Horne-Zeilinger state of $n$ photons in $n$ 1D SRs
coupled to a SQ. In 2015, Hua \emph{et al.} \cite{Hua2} constructed
the fast universal quantum gates on 1D SRs with resonance
operations.

To generate the particular entanglement on MPs, one always needs the
SQ to apply the nonlinear interaction and acts as the coupler
element in previous works. In the processor composed of some
resonators coupled to a SQ \cite{Strauch,Yang,Hua2}, to integrate
more resonators, one should take small coupling strength between
each resonator and the SQ and the large tunable range of the
frequency of the SQ or take the tunable coupling technique between
each resonator and the SQ. Small coupling strength leads to a slow
quantum entangling operation on the SQ and the SR, which limits the
performance of the processor. The large tunable range of the SQ or
the tunable coupling between each resonator and the SQ will increase
the difficulty for the implementation in experiment. On one hand,
the long-coherence-time transmon qubit \cite{Koch}, which is always
used to deal with the quantum computing, can just be tuned
effectively with a frequency range of $\sim 2.5$ GHz
\cite{Schreier}. On the other hand, the tunable coupling is just
realized between a SQ and a SR \cite{Srinivasan}. Moreover, the
length of the SQ ($\sim$ tens of nm) can only deal with the
entangling operation on SRs in a small range, which limits the
ability for the large-scale integration of the processor.

In this letter, we present a scalable device for generating
entanglement on microwave photons in distant  SRs  by using another
common SR $R$ as the quantum bus. In this device,  multiple SRs
$r_j$ (act as quantum information carriers) can be coupled to the
bus in its different positions simply, assisted by superconducting
quantum interferometer devices (SQUIDs). It does not require SQs,
far different from previous quantum processors on MPs. As a critical
task for quantum communication, we give a one-step implementation
scheme to generate the Bell state on two distant SRs $r_1$ and $r_2$
with a high fidelity. This scheme works in the dispersive regime of
$r_j$ and $R$ ($g_j/\Delta_j\ll 1$) and it can be extended to
generate multi-Bell states on different pairs of resonators
simultaneously. Here and after, $g_j$ is the coupling strength
between $r_j$ and $R$. $\Delta_j=\omega_j-\omega_{_R}$ in which
$\omega_j$ and $\omega_R$ are the frequencies of $r_j$ and $R$,
respectively. By using the technique for catching and releasing a MP
state in a SR, our device can work as a Bell-state generator or a
node for long-distance quantum communication.

\begin{figure}[h]       % fig1
%[tpb]
%\par
\begin{center}
\includegraphics[width=8.0cm,angle=0]{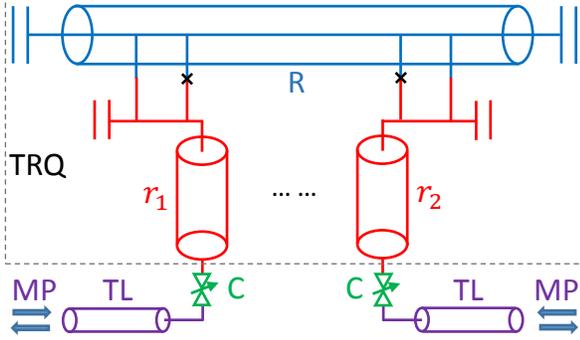}
\end{center}
\caption{(Color online) The scalable quantum-bus-based device for
generating the entanglement on microwave photons (MPs) in different
superconducting resonators.  Here the resonator $r_j$ can catch and
release a microwave photon from or to the transmission line (TL) by
turning on the coupling $C$. The resonator $R$ acts as a quantum
bus. The subsystem composed of three resonators (TRQ) shown in the
dashed-line box is the setup for producing  Bell states on microwave
photons in the two distant resonators $r_1$ and $r_2$.} \label{fig1}
\end{figure}

Our device for generating  Bell states on microwave photons in
distant SRs is composed of some 1D SRs $r_j$ ($j=$ $1$, $2$
$\cdots$) coupled to a quantum bus $R$ assisted by the SQUIDs, shown
in Fig.\ref{fig1}. $r_j$ are placed far enough to each other to
avoid the mutual capacitances and mutual inductive interactions
among them. Quantum information on $|0\rangle_j$ and $|1\rangle_j$
in resonators $r_j$ can be released and catched by turning on the
coupler $C$. To show the principle of our device, one just needs to
consider the three-resonator-qubit (TRQ) subsystem composed of
$r_1$, $r_2$, and $R$, shown in the dashed-line box in Fig.
\ref{fig1}. Here, the coupling $C$ should be turned off. In the
interaction picture, the Hamiltonian of the TRQ subsystem can be
expressed as ($\hbar=1$, under the rotating-wave approximation)
\cite{Peropadre}
\begin{eqnarray}              %eq1
H  &=& \sum_j g_j(a^{+}b_j e^{-i\Delta_jt} +
ab_j^{+}e^{i\Delta_jt}). \label{all}
\end{eqnarray}
Here, $g_j$ is decided by the coupling position on $R$ and can be
tuned by an external flux field through the SQUID \cite{Peropadre}.
$a^{+}$ and $b_j^{+}$ are the creation operators of the resonators
$R$ and $r_j$, respectively.

In order to generate the Bell state on $r_1$ and $r_2$ in the TRQ
subsystem, let us consider the dispersive regime of $r_j$ and $R$.
In the Schr\"odinger picture, Hamiltonian $H$ can be rewritten as
\begin{eqnarray}              %eq2
H'  &=& \omega_{_R} a^{+}a +\omega_1 b_1^{+}b_1 +\omega_2 b_2^{+}b_2 \nonumber\\
&& +\; g_1(a^{+}b_1 + ab_1^{+}) + g_2(a^{+}b_2 + ab_2^{+}).
\label{effect1}
\end{eqnarray}
By taking the unitary transformation \cite{Blais}
\begin{eqnarray}              %eq3
U = \exp \left[\frac{g_1}{\Delta_1}(ab_1^{+}-a^{+}b_1) +
\frac{g_2}{\Delta_2}(ab_2^{+}-a^{+}b_2) \right] \label{effect2}
\end{eqnarray}
to the second order in the small parameters $g_j /\Delta_j$, one can
get
\begin{eqnarray}              %eq4
H'' = UH'U^{+} %\nonumber\\
 &\approx &  \omega_{_R}'a^{+}a + \omega_1'b_1^{+}b_1+\omega_2'b_2^{+}b_2 \nonumber\\
&& +\;
g'(b_1b_2^{+}+b_1^{+}b_2).\;\;\;\;\;\;\;\;\;\;\;\;\;\;\;\;\;\;\;
\label{effect3}
\end{eqnarray}
Here, $g'=\frac{g_1 g_2}{\Delta_1 \Delta_2}(\Delta_1+\Delta_2)$,
$\omega_{_R}'=\omega_{_R}-\chi_1-\chi_2$,
$\omega_1'=\omega_1+\chi_1$, and $\omega_2'=\omega_2+\chi_2$
($\chi_j=\frac{g_j}{\Delta_j}$). In the interaction picture, the
Hamiltonian $H''$ is rewritten as
\begin{eqnarray}              %eq5
H_{eff}  &=& g'(b_1^{+}b_2e^{-i\delta t} + b_1b_2^{+}e^{i\delta t}),
\label{effect4}
\end{eqnarray}
in which the detuning value $\delta= \omega_1'-\omega_2'$. In the
subspace composed of $|0\rangle_1|1\rangle_2$ and
$|1\rangle_1|0\rangle_2$, the evolution of the state $|
\psi(t)\rangle$  of  the TRQ subsystem can be solved with the
equation of motion
\begin{eqnarray}           %  eq6
i \frac{\partial|\psi(t)\rangle}{\partial t}= H_{eff}
|\psi(t)\rangle. \label{1}
\end{eqnarray}
Here, $| \psi(t)\rangle$ is the linear combination of the states
$|0\rangle_1 |1\rangle_2$ and $|1\rangle_1 |0\rangle_2$, that is,
\begin{eqnarray}           %  eq7
|\psi(t)\rangle= c_{_{0,1}}(t)|0\rangle_1
|1\rangle_2+c_{_{1,0}}(t)|1\rangle_2 |0\rangle_1 \label{2}
\end{eqnarray}
in which $c_{_{0,1}}(t)$ and $c_{_{1,0}}(t)$ can be expressed as
\begin{eqnarray}            %  eq8
\frac{\partial c_{_{1,0}}}{\partial t} &=&-i g' c_{_{0,1}}e^{i\delta t}, \nonumber\\
\frac{\partial c_{_{0,1}}}{\partial t} &=&-i g'
c_{_{1,0}}e^{-i\delta t}. \label{3}
\end{eqnarray}
The general solution for $c_{_{0,1}}(t)$ and $c_{_{1,0}}(t)$ is
\cite{Nielsen}
\begin{eqnarray}           %  eq9
c_{_{0,1}}(t) &=&\{c_{_{0,1}}(0)\left[\cos \left(\frac{\Omega
t}{2}\right)+\frac{i\delta }{\Omega }
\sin \left(\frac{\Omega t}{2}\right)\right] \nonumber\\
&& -\frac{2ig'}{\Omega }c_{_{1,0}}(0)\sin \left(\frac{\Omega t
}{2}\right)\}e^{i-\delta t/2}, \nonumber\\
c_{_{1,0}}(t) &=&\{c_{_{1,0}}(0)\left[\cos \left(\frac{\Omega
t}{2}\right)-\frac{i\delta }{\Omega }
\sin \left(\frac{\Omega t}{2}\right)\right] \nonumber\\
&& -\frac{2ig'}{\Omega }c_{_{0,1}}(0)\sin \left(\frac{\Omega t
}{2}\right)\}e^{i\delta t/2}. \label{4}
\end{eqnarray}
Here $\Omega ^{2}=4g'^{2}+\delta^{2}$.

Suppose the initial state of the TRQ subsystem is
$|\Psi_1\rangle=|1\rangle_1|0\rangle_2|0\rangle_{_R}$ which
indicates that there is an MP in the resonator $r_1$. In the
resonance regime of $r_1$ and $r_2$ (i.e., $\delta=0$), the state of
the TRQ subsystem evolves to
\begin{eqnarray}              %eq10
\Psi (t) = C_1 (t) |\Psi_1\rangle + C_2 (t) |\Psi_2\rangle
\label{effect}
\end{eqnarray}
at the time $t$ with the coefficients
\begin{eqnarray}          %eq11
C_1 (t) &=& \cos{g't},
\label{c1}\\
C_2 (t) &=&  -i\sin{g't}. \label{c2}
\end{eqnarray}
Here $|\Psi_2\rangle=|0\rangle_1|1\rangle_2|0\rangle_{_R}$. Under
the condition $\omega_j<\omega_{_R}$ and after an operation time of
$g't=\frac{\pi}{4}$, one will get the final state
\begin{eqnarray}              %eq13
|\Psi_{+}^{1,2}\rangle=\frac{1}{\sqrt{2}}(|0\rangle_1|1\rangle_2+i|1\rangle_1|0\rangle_2).
\label{EPR1}
\end{eqnarray}
By taking $g't=\frac{\pi}{4}$ and $\omega_j>\omega_{_R}$, one can
get the state
\begin{eqnarray}              %eq14
|\Psi_{-}^{1,2}\rangle=\frac{1}{\sqrt{2}}(|0\rangle_1|1\rangle_2-i|1\rangle_1|0\rangle_2).
\label{EPR2}
\end{eqnarray}
Both $|\Psi_{+}^{1,2}\rangle$ and $|\Psi_{-}^{1,2}\rangle$ are the
Bell states on microwave photons in the two distant resonators $r_1$
and $r_2$.

\begin{figure}%[h]       % fig2
%[tpb]
%\par
\begin{center}
\includegraphics[width=8cm,angle=0]{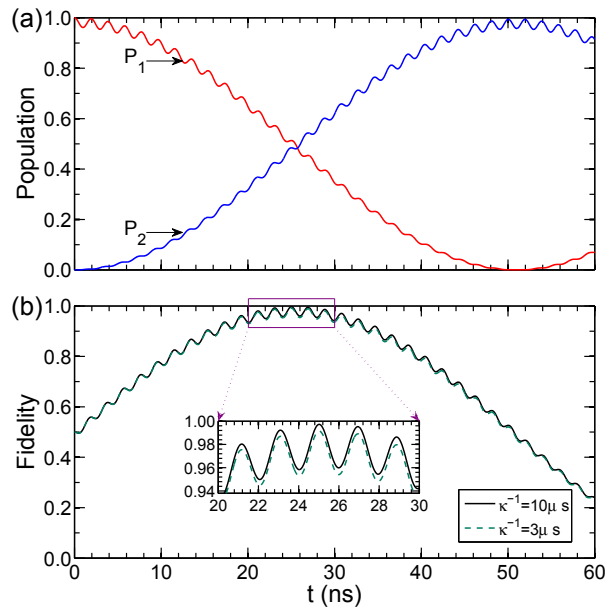}
\end{center}
\caption{(Color online) (a) The populations of a MP in $r_1$ and
$r_2$. $P_1$ and $P_2$ with red and blue solid lines represent the
populations of a MP in $r_1$ and $r_2$, respectively. (b) The
fidelity of the Bell state $|\Psi_{+}^{1,2}\rangle$ vs the operation
time $t$ and different decay rates of the resonators. The black
solid line and the dark green dashed line represent the fidelities
with the decay rates $\kappa^{-1}=10$ $\mu$s and $\kappa^{-1}=3$
$\mu$s, respectively.} \label{fig2}
\end{figure}

To show the feasibility of our one-step scheme for generating the
Bell state $|\Psi_{+}^{1,2}\rangle$, we numerically simulate its
fidelity with the feasible parameters. The evolution of the TRQ
subsystem can be described by the master equation \cite{Blais}
\begin{eqnarray}              %eq15
\frac{d\rho }{dt} \!=\! -i[H,\rho ]+\kappa_{_R} D[a]\rho
+\kappa_{_1} D[b_1]\rho+\kappa_{_2} D[b_2]\rho.
\label{masterequation}
\end{eqnarray}
Here, $\kappa_n$ ($n=R,1,2$) is the decay rate of the resonator
$r_n$. $D[L]\rho=(2L\rho L^{+}-L^{+}L\rho-\rho L^{+}L)/2$. $\rho$ is
the realistic density operator after our Bell-state generation
scheme with the initial state $|\Psi_1\rangle$ and the realistic
Hamiltonian $H$. The influence from the SQUIDs has been neglected as
their frequencies  are designed to detune with those of the
resonators largely \cite{Peropadre}. By taking the parameters as:
$\omega_1/(2\pi)=\omega_2/(2\pi)=5.75$ GHz,
$\omega_{_R}/(2\pi)=6.25$ GHz, $g_1/(2\pi)=g_2/(2\pi)=50$ MHz, and
$\kappa_{_R}^{-1}=\kappa_1^{-1}=\kappa_2^{-1}=\kappa^{-1}=10$
$\mu$s, we numerically simulate  the populations of a microwave
photon in resonators $r_1$ and $r_2$ varying  with  the operation
time, shown in Fig.\ref{fig2}  (a). The definition of the population
is
\begin{eqnarray}        %  eq16
P_m &=& \langle \Psi_m|\rho(t)|\Psi_m\rangle. \label{p}
\end{eqnarray}
Here $m=$ $1$, $2$. The fidelity of the state
$|\Psi_{+}^{1,2}\rangle$ generated with the initial state
$|\Psi_1\rangle=|1\rangle_1|0\rangle_R|0\rangle_2$ in the TRQ
subsystem varying with the operation time is simulated by using the
definition
\begin{eqnarray}        %  eq17
F &=& \langle \Psi_{+}^{1,2}|\rho(t)|\Psi_{+}^{1,2}\rangle \label{F}
\end{eqnarray}
and it is shown in Fig.\ref{fig2} (b), in which the fidelity varies
with the operation time and different decay rates of resonators
(those with $\kappa^{-1}=10$ $\mu$s and $\kappa^{-1}=3$ $\mu$s are
plotted with the black solid line and the dark green dashed line,
respectively). Here the maximal fidelity with the decay rate of
$\kappa^{-1}=10$  $\mu$s can reach  $99.73\%$ within about $25$ ns
and the one with the decay rate of $\kappa^{-1}=3$ $\mu$s can reach
$99.1\%$. It is worth noticing that the best quality factor of a 1D
superconducting resonator has reached $Q \sim 2 \times 10^6$
\cite{Megrant}, which indicates the life time of a microwave photon
in the resonator is about $50$ $\mu$s with considering the relation
$\kappa=\omega_r/Q$ \cite{Blais} ($\omega_r$ is the frequency of the
resonator). The coupling strengths $g_1$ and $g_2$ taken here are
the theoretic result discussed in Ref.\cite{Peropadre} which fulfill
the rotating-wave approximation in Eq.(\ref{all}).

\begin{figure}%[h]       % fig3
%[tpb]
%\par
\begin{center}
\includegraphics[width=7.0cm,angle=0]{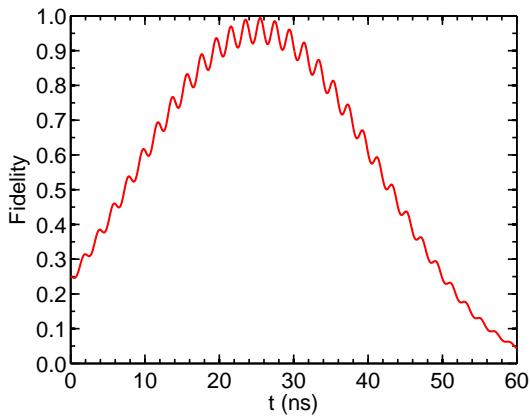}
\end{center}
\caption{(Color online) The fidelity of the multi-Bell state $\vert
\Psi\rangle$ vs the operation time $t$.} \label{fig3}
\end{figure}

More resonators $r_j$ ($j=$ $1$, $2$, $3$, $4$ $...$) can be coupled
to the quantum bus $R$ simply in our device for  generating
multi-Bell states on different resonator pairs simultaneously in the
dispersive regime of $r_j$ and $R$. Here, we take the generation of
two Bell states on two resonator pairs as an example to describe its
principle. In detail, the five-resonator-qubit (FRQ) subsystem
composed of $r_1$, $r_2$, $r_3$, $r_4$, and $R$ is used to generate
two pairs of Bell states on MPs.  Similar to $r_1$ and $r_2$ shown
in Fig.\ref{fig1}, $r_3$ and $r_4$ are coupled to the bus $R$ and
they are not drawn. In the dispersive regime of $r_j$ and $R$, $r_j$
are detuned with $R$ largely. Meanwhile, if one takes the
frequencies of $r_1$ and $r_2$ to detune with those of $r_3$ and
$r_4$ largely, our one-step scheme for the generation of the Bell
state on $r_1$ and $r_2$ will works independently with the one on
$r_3$ and $r_4$. Moreover, $g'$ of $r_1$ and $r_2$ should be
equivalent to the one of $r_3$ and $r_4$ for generating the Bell
states simultaneously in the FRQ subsystem.  We numerically simulate
the fidelity $F^{1,2,3,4}$ for generating the state
$\frac{1}{\sqrt{2}}(|0\rangle_1|1\rangle_2+i|1\rangle_1|0\rangle_2)$
on $r_1$ and $r_2$ and the state
$\frac{1}{\sqrt{2}}(|0\rangle_3|1\rangle_4-i|1\rangle_3|0\rangle_4)$
on $r_3$ and $r_4$ simultaneously in this FRQ subsystem. The
parameters of $r_3$ and $r_4$ are taken as
$\omega_3/(2\pi)=\omega_4/(2\pi)=6.75$ GHz,
$g_3/(2\pi)=g_4/(2\pi)=50$ MHz, and $\kappa_3^{-1}=\kappa_4^{-1}=10$
$\mu$s. The parameters of $r_1$, $r_2$, and $R$ are the same as the
ones for generating the state
$\frac{1}{\sqrt{2}}(|0\rangle_1|1\rangle_2+i|1\rangle_1|0\rangle_2)$
on $r_1$ and $r_2$ in our TRQ subsystem. The definition of the
fidelity is
\begin{eqnarray}        %  eq18
F^{1,2,3,4} &=& \langle \Psi|\rho'(t)|\Psi\rangle \label{F1}
\end{eqnarray}
in which $\rho'(t)$ is the realistic density operator of the FRQ
subsystem with the initial state
$|1\rangle_1|0\rangle_2|1\rangle_3|0\rangle_4|0\rangle_{_R}$. $\vert
\Psi\rangle
=\frac{1}{2}(|1\rangle_1|0\rangle_2+i|0\rangle_1|1\rangle_2)\otimes(|1\rangle_3|0\rangle_4-i|0\rangle_3|1\rangle_4)
\otimes |0\rangle_{_R}$. The result of our simulation for the
fidelity of the final state is  shown in Fig. \ref{fig3}. It can
reach $99.5\%$ within about $25.5$ ns. $g_j/\Delta_j$ taken here is
not small enough to ignore the influence from the interactions
between $r_j$ and $R$. That is, under the same parameters of $r_1$,
$r_2$, and $R$ in the TRQ subsystem, the maximal fidelity of the
state
$\frac{1}{\sqrt{2}}(|1\rangle_1|0\rangle_2+i|0\rangle_1|1\rangle_2)$
on $r_1$ and $r_2$ achieved in the FRQ subsystem is smaller than the
one in the TRQ subsystem, and the fidelity is also reduced faster
than the one in the TRQ subsystem.

Although our schemes work in the dispersive regime of $r_j$ and $R$,
one can insert a SQUID in each resonator $r_j$ to tune their
frequencies \cite{Sandberg} to make $r_j$ to resonate with $R$ and
make the frequencies of MPs equivalent to each other before they are
released to the TL. In our scheme, the MPs in the initial states
$|\Psi_1\rangle$ and
$|1\rangle_1|0\rangle_2|1\rangle_3|0\rangle_4|0\rangle_{_R}$ can be
caught from the TLs. The oscillations with small amplitudes of the
populations and the fidelities in Fig. \ref{fig2} and Fig.
\ref{fig3} come from the interactions between $r_j$ and $R$, which
can be removed by increasing the detuning value $\Delta_j$ or reduce
the coupling strength $g_j$. To generate more Bell states on
different resonator pairs, one should also remove the influence from
the bus $R$ effectively and take a  smaller $g_i/\Delta_j$ to reduce
the indirect interaction among more resonator pairs with difference
frequencies.

In summary, we have proposed a scalable  quantum-bus-based device
for generating entangled states on microwave photons. It is composed
of some resonators $r_j$ coupled to a quantum bus $R$ assisted by
SQUIDS. With this  device, we present a high-fidelity one-step
scheme for the generation of the Bell state on the two distant
resonators $r_1$ and $r_2$. The scheme works in the dispersive
regime of $r_j$ and $R$. Besides, in the dispersive regime, the
scheme can also be extended to generate the multi-pair Bell states
one different pairs of SRs $r_j$ simultaneously and we take the case
for generating two Bell states on two pairs of resonators
simultaneously as an example to describe its principle, which can be
achieved with a high fidelity as well.

%\bigskip

This work was supported by the National Natural Science Foundation
of China under Grant Nos. 11174039 and 11474026, and the Fundamental
Research Funds for the Central Universities under Grant No.
2015KJJCA01.

\end{document}